\documentclass[fleqn,twoside]{article}
\usepackage{espcrc2}

\usepackage{graphicx}
\usepackage{epsfig}
\usepackage[figuresright]{rotating}

\newcommand{\AmS}{{\protect\the\textfont2
  A\kern-.1667em\lower.5ex\hbox{M}\kern-.125emS}}
\newcommand{\be}{\begin{equation}}
\newcommand{\ee}{\end{equation}}
\newcommand{\ben}{\begin{eqnarray}}
\newcommand{\een}{\end{eqnarray}}
\newcommand{\bra}{\langle}
\newcommand{\ket}{\rangle}

\title{Simulation of Wilson fermion at finite isospin density\thanks{Presented by T.Takaishi} 
       }

\author{A. Nakamura\address[MCSD]{RIISE, Hiroshima University,  Higashi-Hiroshima 739-8521, Japan}%
        ~and   
        T. Takaishi\address{Hiroshima University of Economics, Hiroshima 731-0192, Japan }
}
       
\begin{document}

\begin{abstract}
For QCD with Wilson fermions at isospin chemical potential 
we study the finite phase transition on an $8^3\times4$ lattice at $\kappa=0.15$.
We use two gauge actions: Wilson action and DBW2 action.
Both actions give the same results.
The phase diagram is qualitatively similar to the one obtained for QCD at small baryon chemical potential.
We also calculate the number density for various isospin chemical potentials.
\end{abstract}

\maketitle

\section{INTRODUCTION}

Lattice simulations of finite density QCD
are difficult due to the sign problem.
Namely the action is complex and the complex phase 
fluctuates, which makes the simulations difficult.
Recently considerable progress has been 
made for small baryon chemical potential $\mu_B$\cite{RECENT}. 
There exist several approaches to small $\mu_B$:
Taylor expansion method\cite{QCDTARO},
reweighting method\cite{RE} 
and imaginary chemical potential method\cite{IMG}.

Apart from the complex action, 
a model with a positive measure like isospin model 
can be used to obtain insights to QCD at finite $\mu_B$.
Moreover it might be expected that the phase diagram of QCD at small $\mu_B$ is 
similar to that at small isospin chemical potential $\mu_I$\cite{KOKUT}.  

In this study we use Wilson fermions with  $\mu_I$ 
and study the phase diagram and the number density.

\section{ISOSPIN DENSITY}

Lattice QCD partition function with $N_f$ flavors is given by
\be
Z=\int [dU](\prod_i^{N_f} \det D(\mu_i)) \exp(S_g[U])
\ee
where
\be
S_g[U]=\frac{\beta}3((1-8c_1) \sum {\sf Re Tr} W11  +c_1 \sum {\sf ReTr} W12 )
\label{eq:gauge}
\ee
and $D(\mu_i)$ is the Wilson fermion matrix at $\mu_i$.
$W11$ and $W12$ stand for $1\times1$ 
and $1\times2$ loops respectively.
For $N_f=2$ with $\mu_u=-\mu_d\equiv\mu_I$
we obtain
\ben
\det D(\mu_u)\det D(\mu_d) & = & \det D(\mu_I)\det D(-\mu_I) \nonumber \\
                       & = & |\det D(\mu_I)|^2
\een
where a relation $D(\mu_i)=\gamma_5 D^\dagger(-\mu_i)\gamma_5$ is used.
We call $\mu_I$ isospin chemical potential.
For this $\mu_I$ the measure is positive definite and
the standard Monte Calro technique can be applied.

The parameter $c_1$ in eq.(\ref{eq:gauge})
is the one which specifies the gauge action. For example
$c_1=0$ for the Wilson gauge action, and
$c_1=-1.4089$ for the DBW2 action\cite{DBW2}.

\section{SIMULATIONS}

We generate configurations on an $8^3\times 4$ lattice 
at $\kappa=0.15$ by the hybrid Monte Calro (HMC) algorithm.
We use the Wilson gauge action and the DBW2 action\cite{DBW2}.
Simulation parameters are summarized in Tables 1 and 2.
We choose a step size so that the acceptance of 
the HMC algorithm becomes $60 \sim 70$\% which 
gives the maximum performance of the HMC algorithm with
the 2nd order leapfrog integrator\cite{HOHMC}.
This optimum acceptance $\approx 60 \sim 70$\%
does not depend on the details of the action 
we take. It depends on the order of the leapfrog integrator.

\begin{table}[t]
\begin{tabular}{clccc} \hline
 $\mu_I$  & $\beta$ & $\Delta t$ & Acc. & Traj. \\ \hline
0.0     &  5.38   & 1/14 & 0.667(4) & 13000  \\
0.1     &  5.37   & 1/14 & 0.667(5) & 12000  \\
0.2     &  5.365  & 1/14 & 0.652(5) & 15500  \\
0.3     &  5.34   & 1/14 & 0.647(5) & 12500 \\ \hline
\end{tabular}
\caption{Simulation parameters for the Wilson gauge action.
$\Delta t$ is the step size and the trajectory length is
set to 1. "Acc." and "Traj." stand for Acceptance and \# of trajectories
respectively.
}
\vspace{-2mm}
\end{table}

\begin{table}
\vspace{-3mm}
\begin{tabular}{clccc} \hline
 $\mu_I$  & $\beta$ & $\Delta t$ & Acc. & Traj. \\ \hline
0.0     &  0.68   & 1/16 & 0.644(3) & 28000  \\
0.1     &  0.67   & 1/16 & 0.646(2) & 17800  \\
0.2     &  0.66   & 1/16 & 0.653(3) & 30000  \\ \hline
\end{tabular}
\caption{Same as in Table 1 but for the DBW2 action.}
\vspace{-4mm}
\end{table}

\begin{figure}[t]
\begin{center}
\epsfig{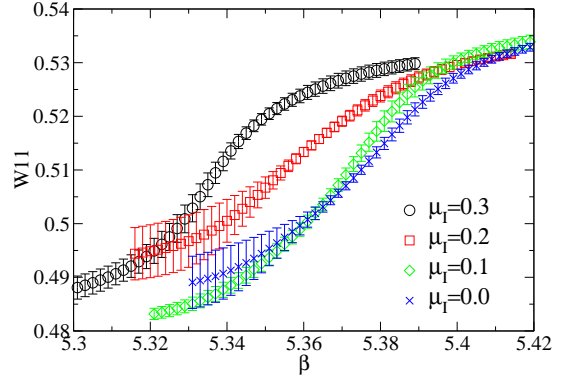}
\vspace{-0.8cm}
\caption{\label{fig:plaq-beta}
$1\times1$ Wilson loop from the Wilson gauge action 
for various $\mu_I$
as a function of $\beta$.
}
\end{center}
\vspace{-0.2cm}
\end{figure}

\section{RESULTS}

In order to locate the phase transition point 
we measure susceptibility for various observables.
Since for each $\mu_I$ we performed a single simulation at one $\beta$,
the reweighting method\cite{SWENDSEN} was used to investigate 
a region in the vicinity of the $\beta$. 
The expectation value of an observable $O$ at $\beta^\prime$ 
can be obtained through a single simulation at $\beta$ by
\be
\bra O\ket_{\beta^\prime} =\frac{\bra O \exp(\Delta \beta)\ket_\beta }
{\bra \exp(\Delta \beta \ket_\beta}
\ee
where $\Delta \beta=\beta^\prime-\beta$.

Measurements are done for $1\times1$, $1\times2$, $2\times2$ Wilson and Polyakov loops.
Figure \ref{fig:plaq-beta} shows the $1\times1$ loop from 
the Wilson gauge action for various $\mu_I$.
The critical coupling constant $\beta_c$ decreases as $\mu_I$ increases.
The precise value of $\beta_c$ is estimated by measuring susceptibilities of 
the observables.


\begin{figure}[t]
\begin{center}
\epsfig{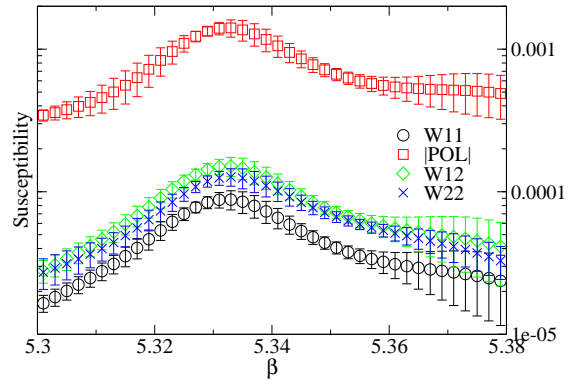}
\vspace{-0.9cm}
\caption{\label{fig:susb534}
Susceptibility from various observables at $\mu_I=0.3$.
The Wilson gauge action is used.
}
\end{center}
\vspace{-4mm}
\end{figure}

\begin{figure}[t]
\begin{center}
\epsfig{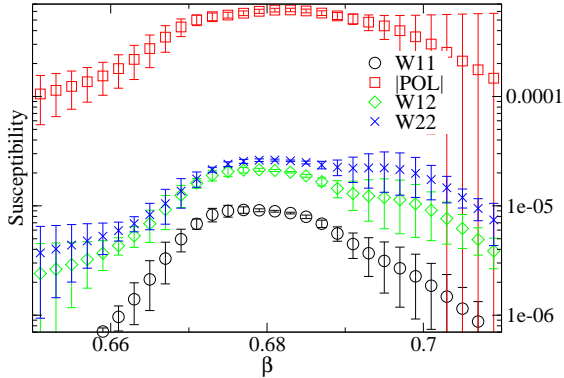}
\vspace{-0.9cm}
\caption{\label{fig:susb068}
Same as in Figure \ref{fig:susb534} but from the DBW2 action at $\mu_I=0.0$.
}
\end{center}
\vspace{-4mm}
\end{figure}

\begin{figure}[t]
\begin{center}
\epsfig{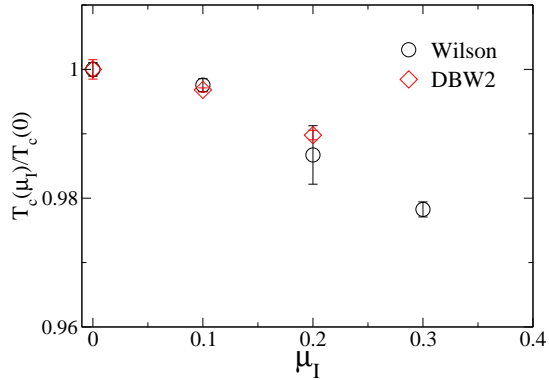}
\vspace{-9mm}
\caption{\label{fig:tcvsmu2}
Phase diagram in the $T-\mu_I$ plane.
$T$ is normalized by $T_C$ at $\mu_I=0.0$.
}
\end{center}
\vspace{-6mm}
\end{figure}

Typical examples of susceptibilities from various observables are shown 
in Figures \ref{fig:susb534}-\ref{fig:susb068}.
All susceptibilities from different observables give similar $\beta_c$.
From $\beta_c$ (Wilson gauge action) evaluated from the $1\times 1$ loop susceptibility we obtain 
\be
\beta_c(\mu_I)=5.38-0.54\mu_I^2.
\ee 
The coefficient of $O(\mu_I^2)$ term is 
similar to those of \cite{IMG}. 
Figure \ref{fig:tcvsmu2} shows the phase diagram in the $T-\mu_I$ plane. 
$\beta$ is converted to $T$ by the 2 loop $\beta$ function. 
Two phase diagrams from the Wilson and DBW2 gauge actions 
are in good agreement. 
The phase diagrams are qualitatively similar 
to the one obtained with KS fermions at small $\mu_B$\cite{RE,IMG}.

\begin{figure}
\begin{center}
\epsfig{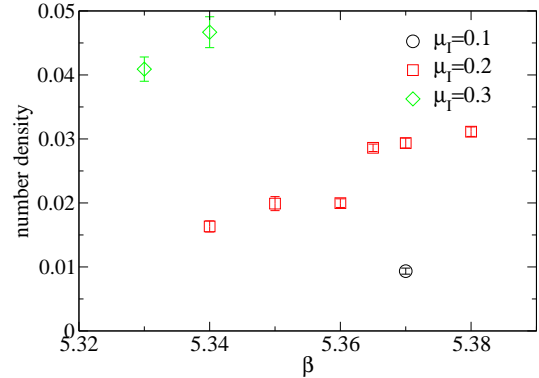}
\vspace{-0.9cm}
\caption{\label{fig:number}
Number density for various $\mu_I$ as a function of $\beta$.
We generated $25-50$ configurations at each $\beta$.
}
\end{center}
\vspace{-7mm}
\end{figure}

We also calculate the number density $n_d$ defined by
$\displaystyle
\frac1V Tr\frac1D \frac{\partial D}{\partial \mu_I}.
$
Figure \ref{fig:number} shows $n_d$ for different $\mu_I$.
$n_d$ seems to increases with $\beta$.

\section{DISCUSSION}
We have studied QCD at finite isospin density with Wilson fermions.
The results show that the phase diagram is qualitatively similar to the one 
obtained  at small $\mu_B$.
The Wilson and DBW2 gauge actions give the same phase diagram.
More quantitative analysis is needed to confirm the present results.

\vspace{0cm}
\section*{ACKNOWLEDGEMENTS}
The simulations were performed on NEC SX-5 at RCNP, Osaka University and
at Yukawa Institute, Kyoto University.

\end{document}